\documentclass[epsf,12pt]{article}
\input psfig.sty
\newcommand{\beq}[2]{\begin{equation}#1\label{#2}\end{equation}}
\newcommand{\ceq}[1]{(\ref{#1})}
\newcommand{\mbd}[1]{\mbox{\bf #1}}

\newcommand{\bx}{\mbd{x}}

\newcommand{\ba}{\mbd{a}}
\newcommand{\bb}{\mbd{b}}

\newcommand{\bj}{\mbd{j}}
\newcommand{\bJ}{\mbd{J}}

\newcommand{\bS}{\mbd{S}}

\newfont{\mbld}{cmbx10 scaled 800}
\newfont{\cab}{cmsy10 scaled 1200}
\newfont{\scab}{cmsy10 scaled 1000}
\newcommand{\bcas}{\mbox{\cab S}}
\newcommand{\sbcas}{\mbox{\scab S}}

\newcommand{\sbx}{\mbox{\mbld{x}}}

\newcommand{\BF}[1]{\mbox{\boldmath $#1$}}          
\newcommand{\nablab}{\BF{\nabla}}
\begin{document}
\title{A Topological Field Theory with a Finite Number of Connected
Feynman Diagrams}
\author{Franco Ferrari\\
{\it Institute of Physics, University of Szczecin, ul. Wielkopolska 15,}\\
{\it 70-451 Szczecin, Poland}\thanks{e-mail:
fferrari@univ.szczecin.pl}.}


\maketitle

\abstract{
A new topological field theory is constructed, which
 is characterized by cubic interactions similar to those of
non-abelian Chern-Simons field theories, but still retains the
simplicity of the abelian case. The perturbative expansion of this
theory contains in fact only two connected Feynman diagrams, the
propagator and a three vertex. Apart from the Gauss linking number, the
Wilson loop amplitudes generate a further topological invariant, whose
physical and mathematical meaning is investigated.
}
\vfill\eject
\pagestyle{plain}
\section{Foreword}
In several situations it has been experimentally observed that the
topological properties of certain physical systems may influence their
behavior to a relevant extent. This is for instance the case of vortex
structures in nematic liquid crystals \cite{nemliq}
and in $^3He$ superfluids \cite{supflu}. Other
examples are provided by polymers \cite{wasserman}
 or by the lowest lying excitations of
two-dimensional electron gases, which have topological non-trivial
configurations at some filling fractions \cite{quahaleff}.
In the investigation of phenomena
related to the presence of topological constraints in the system, the
use of
quantum or statistical mechanical models
coupled to abelian Chern--Simons (C-S) field
theories \cite{chesim}
has been particularly successful.
One reason of this success is the fact that abelian models
do not require a complex mathematical
treatment as their non-abelian counterparts and thus
their physical meaning is more transparent.

Motivated by possible applications in physics,
the aim of
this work is the construction
of a topological field theory
with non-trivial cubic interactions similar to those of
non-abelian C-S field theories, but which still retains
the
simplicity of the abelian case.
As a result of this effort an exactly solvable
 topological field theory is obtained, called hereafter truncated
topological field theory or briefly TTFT, 
which contains only
two connected Feynman diagrams in its perturbative expansion.
The name of the theory is owing to the fact that any further expansion of
the perturbative series, which could in principle generate new 
diagrams, has been truncated by the introduction of suitable constraints.
From the computation of the Wilson loop amplitudes it turns out
that, apart from the Gauss linking number that is already present in
the abelian C-S field theory, the TTFT delivers a further
topological invariant, which can be interpreted as an Hopf term.

The material presented in this paper is divided as follows.
A naive topological field theory
consisting of three BF--models \cite{bfmod} coupled together by
cubic interaction terms is investigated in Section 2. Since this theory
is topological, it is convenient to choose as its observables the
so-called Wilson loops. Unfortunately, after the insertion of the
Wilson loops in the partition function in order to compute their amplitudes,
one observes that the constraints generated by the
longitudinal components of the fields become inconsistent with the
rest of the equations of motion.
This problem is solved in Section 3 by enlarging the gauge group of
the naive model via the addition of suitable topological terms to its
action. In this way a well defined TTFT is obtained, in which the
longitudinal components of the fields are harmless, because they
correspond to pure gauge field configurations and are thus irrelevant.
In Section 4 the Wilson loops amplitudes of the TTFT are computed in
the Lorentz gauge, showing that they contain a single topological
invariant apart from the Gauss linking number.
The physical and mathematical meaning of this invariant is
investigated. Finally, the Conclusions and a possible extension of the TTFT
are presented in Section 5.
%
\section{Problems with Cubic Interactions in Abelian BF--Models}
In the quest for a topological field theory which generates only a finite
number of topological invariants, it is natural to start from the naive
action:
\beq{S=
\int d^3x\left[\frac\kappa{4\pi}
\tilde{\BF\Omega}^i\cdot(\nablab\times{\BF\Omega}^i)+
\Lambda{\BF\Omega}^1\cdot
({\BF\Omega}^2\times{\BF\Omega}^3)+
\bJ_i\cdot\tilde{\BF\Omega}^i\right]}{snaivevector}
For simplicity, $S$ has been defined here on a three dimensional
Euclidean space.
In Eq.~\ceq{snaivevector} $\kappa$ and $\Lambda$ denote
real coupling constants, while
$\tilde{\BF\Omega}^i$ and ${\BF\Omega}^i$, $i=1,2,3$, form a set of
six abelian
vector fields. The $\bJ_i$'s are assumed to be
conserved external currents, i.e. such
that $\nablab\cdot\bJ_i=0$.
Analogous sources coupled to the fields ${\BF\Omega}^i$ have been
omitted for a reason which will be clear below.
Summation over repeated indices is everywhere
understood.
The action $S$ describes a BF--model \cite{bfmod} with the addition
of a cubic interaction term.
In components, Eq.~\ceq{snaivevector} becomes
\beq{S=
\int d^3x\left[\epsilon^{\mu\nu\rho}\left(\frac\kappa{4\pi}
\tilde\Omega^i_\mu\partial_\nu\Omega^i_\rho+
\Lambda\Omega^1_\mu
\Omega^2_\nu\Omega^3_\rho\right)+J^\mu_i\tilde\Omega^i_\mu\right]}{snaive}
where, as a convention, greek letters label space indices, while
roman letters distinguish different vector fields. Finally, 
$\epsilon^{\mu\nu\rho}$ is the Levi-Civita tensor density
defined so that  $\epsilon^{123}=1$.

\begin{figure}[h]
\vspace{1 truein}
\includegraphics{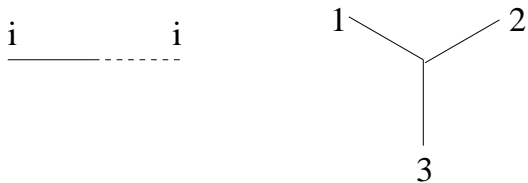}
\vspace{0.2in}
\caption{Feynman rules corresponding to the action \ceq{snaivevector}.
Dashed lines propagate $\tilde{\BF\Omega}^i$ fields while solid lines
are associated to ${\BF\Omega}^i$ fields.}
\label{feyrul}
\end{figure}

Let us note that the fields
$\tilde{\BF\Omega}^i$ play in \ceq{snaivevector}
the role of pure Lagrange multipliers, which
constrain the fields ${\BF\Omega}^i$ and neutralize possible radiative
corrections.
Moreover, at the classical level there are only two
connected Feynman diagrams, which are shown in Fig.~\ref{feyrul}.
They correspond to the field
propagators and to the three-vertex associated to the
cubic interaction term present in
Eq.~\ceq{snaivevector}. Higher order tree diagrams, which  could
in principle be generated by contracting together
the legs of many three-vertices, 
are actually ruled out due to the off-diagonal structure of the
propagators, which forbids any self-interaction among
the fields ${\BF\Omega}^i$.
For the same reason,
the source term $\int d^3x\tilde{\bJ}_i\cdot{\BF\Omega}^i$
for the fields $\tilde{\BF\Omega}^i$,
where the $\tilde{\bJ}_i$'s, $i=1,2,3$, are conserved external currents,
 has been
omitted from Eq.~\ceq{snaivevector}. As a matter of fact, the addition of
such term
would not change the dynamics of the fields ${\BF\Omega}^i$ and, besides,
 it is easy to see that it can be eliminated
by a shift of the Lagrange multipliers $\tilde{\BF\Omega}^i$.

Clearly, the action $S$ is metric independent and its topological properties
are not spoiled by quantum corrections, since the latter vanish identically
as previously remarked. Thus, we are in presence of a topological
field theory, which is also invariant under the following abelian
gauge transformations
\beq{\tilde{\BF\Omega}^i(x)\rightarrow\tilde{\BF\Omega}^i(x)+\nablab
\tilde\lambda^i(x)}
{gftilfie}
As a consequence, one may choose as observables
metric independent and gauge invariant operators like
the Wilson loops:
\beq{W_i(C)={\rm exp} \left[i\oint_\Gamma d\bx\cdot{\BF\Omega}^i\right]}{wldef}
$\Gamma$ is defined here as a superposition of closed non-intersecting
paths $\gamma,\gamma',\gamma'',\ldots$,
i.~e. $\Gamma=\gamma+\gamma'+\gamma''+\ldots$, so that a generic
correlation function of Wilson loops is given
by\footnotemark{}\footnotetext{Actually, we will see later that, due
to the simplicity of the theory under consideration, the only relevant
correlation function of Wilson loops occurs when $\Gamma_i=\gamma_i$,
$i=1,2,3$.}:
\beq{\langle W_1(\Gamma_1) W_2(\Gamma_2) W_3(\Gamma_3)
\rangle=\int\left[\prod_{i=1}^3{\cal D}{\BF\Omega}^i
{\cal D}\tilde{\BF\Omega}^i\right]e^{-iS}}{wlampl}
In writing the above equation, the insertion of Wilson loops has been
taken into account by making the special choice of external currents:
\beq{J_i^\mu=\oint_{\Gamma_i}dx_i^\mu\delta(x-x_i)}{currdef}
$\delta(x)$ being the Dirac $\delta-$function.

Unfortunately, any attempt to compute the amplitude \ceq{wlampl} runs
into troubles due to the
longitudinal components of the fields, whose role has not
been discussed so far. In the case of the $\tilde{\BF\Omega}^i$-fields, it is possible
to get rid of them by fixing the gauge in such a way that:
\beq{\nablab\cdot\tilde{\BF\Omega}^i=0}{otgaufix}
Yet, the undamped
 longitudinal components of the fields ${\BF\Omega}^i$ remain
in the cubic interaction term of Eq.~(\ref{snaivevector}) and introduce
new constraints which, without any treatment, lead to inconsistences
in the theory.
To see how the problem arises, we investigate the
classical equations of motion
associated to the action \ceq{snaivevector}.
A variation of $S$ with respect
to the fields $\tilde\Omega^i_\mu$ produces the constraints:
\beq{\frac\kappa{4\pi}\nablab\times{\BF\Omega}^i+\bJ_i=0}{confromot}
An analogous variation with respect to the fields $\Omega^i_\mu$
yields as a result the following relations:
\begin{eqnarray}
\frac\kappa{4\pi}\nablab\times\tilde{\BF\Omega}^1
& + &\Lambda{\BF\Omega}^2\times
{\BF\Omega}^3=0\label{Aeom}\\
\frac\kappa{4\pi}\nablab\times\tilde{\BF\Omega}^2
& - &\Lambda{\BF\Omega}^1\times
{\BF\Omega}^3=0 \\
\frac\kappa{4\pi}\nablab\times\tilde{\BF\Omega}^3
 & + & \Lambda{\BF\Omega}^1\times
{\BF\Omega}^2=0\label{Ceom}
\end{eqnarray}
It is easy to check that the general solutions of the constraints
\ceq{confromot} are:
\beq{\Omega^i_\mu=b_\mu^i+\partial_\mu\omega^i}{gensolcon}
In the above equation we have put
\beq{b_\mu^i(x)=\frac 1\kappa\epsilon_{\mu\alpha\beta}\int d^3y
\frac{(x-y)^\alpha}{|x-y|^3}J^\beta_i(y)=\frac
1\kappa\epsilon_{\mu\alpha\beta}
\oint_{\Gamma_i}dx_i^\beta\frac{(x-x_i)^\alpha} {|x-x_i|^3}}
{consol}
while the $\omega^i(x)$ represent
differentiable functions, which take into
account the longitudinal components of the vectors $\Omega_\mu^i(x)$.
The form of the $\omega^i(x)$'s cannot be determined from Eqs.~\ceq{confromot}.
At this point, it is possible to solve also Eqs.~(\ref{Aeom}--\ref{Ceom})
exactly with respect to the $\tilde{\BF\Omega}^i$.
However, this is not the end of the story, because there are further
constraints which can be obtained by 
applying the differential operator $\partial_\mu$
to Eqs.~(\ref{Aeom}--\ref{Ceom}). Exploiting
Eqs.~\ceq{confromot} in order to evaluate the curls of the
${\BF\Omega}^i$--fields, one finds the following  relations:
\begin{eqnarray}
{\BF\Omega}^1\cdot\bJ_2-{\BF\Omega}^2\cdot\bJ_1&=&0\label{Aextcon}\\
{\BF\Omega}^2\cdot\bJ_3-{\BF\Omega}^3\cdot\bJ_2&=&0\label{Bextcon}\\
{\BF\Omega}^3\cdot\bJ_1-{\BF\Omega}^1\cdot\bJ_3&=&0\label{Cextcon}
\end{eqnarray}
The solutions of Eqs.~(\ref{Aextcon}--\ref{Cextcon}):
\beq{{\BF\Omega}^i=\bJ_i}{solextcon}
are inconsistent with Eqs.~\ceq{confromot} if $\bJ_i\ne0$.
In fact, since the currents $\bJ_i$ are
conserved by assumption, Eqs.~\ceq{consol} and \ceq{solextcon}
 give two different and clearly incompatible
expressions for the transverse components of the ${\BF\Omega}^i$ fields.

In the next Section it will be shown how to solve this problem.
\section{Solving The Problems: The Truncated Topological Field Theory}
We have seen in the previous Section that the abelian
BF--model with a cubic interaction term defined in Eq.~\ceq{snaivevector}
is inconsistent if the external currents $\bJ_i$ are different from zero.
The difficulties come from the undamped longitudinal components
of the ${\BF\Omega}^i$--fields. In fact, these components
are responsible for the
constraints (\ref{Aextcon}--\ref{Cextcon}), which are incompatible
with the other classical
equations of motion of the theory.
A possible strategy to overcome this problem is to
add suitable terms
to the action $S$, so that the new action is
invariant with respect to gauge
transformations of the kind:
\beq{{\BF\Omega}^i(x)\longrightarrow{\BF\Omega}^i(x)+\nablab\lambda^i(x)}
{gautraome}
The idea behind this strategy is that,
once gauge invariance is established, the longitudinal components
of the ${\BF\Omega}^i$'s become irrelevant. For instance,
they may be easily eliminated
choosing 
a gauge
condition in which the  fields are purely transverse.

To start with, we compute first of all the variations
$\delta_{\lambda^i}S$ of the action $S$ under the gauge
transformations \ceq{gautraome}.
After a few calculations one finds:
\begin{eqnarray}
\delta_{\lambda^1}S&=&\Lambda\int d^3x\nablab\lambda^1\cdot
({\BF\Omega}^2\times{\BF\Omega}^3)\label{Afvar}\\
\delta_{\lambda^2}S&=&-\Lambda\int d^3x\nablab\lambda^2\cdot
({\BF\Omega}^1\times{\BF\Omega}^3)\label{Bfvar}\\
\delta_{\lambda^3}S&=&\Lambda\int d^3x\nablab\lambda^3\cdot
({\BF\Omega}^1\times{\BF\Omega}^2)\label{Cfvar}
\end{eqnarray}
In deriving the above equations
it has not been
taken into account
the fact that there is no real dynamics
in our theory and, for this reason, the expression of the variations
$\delta_{\lambda^i}S$ is unnecessarily complicated. Indeed, in the
right hand sides of 
Eqs.~(\ref{Afvar}--\ref{Cfvar})
there is a
linear dependence
 on the fields ${\BF\Omega}^i$ which is hidden.
To show that, we remember that
the transverse components
of the ${\BF\Omega}^i$'s are bounded to live in the
subspace of all classical field configurations determined by the constraints
\ceq{confromot} and coincide with the
vectors $\bb^i$ given in Eq.~\ceq{consol}.
Exploiting these constraints, it is possible to rewrite
the variations $\delta_{\lambda^i}S$ in the form:
\begin{eqnarray}
\delta_{\lambda^1}S&=&\frac{4\pi\Lambda}\kappa\int d^3x
\lambda^1\left[
\bJ_2\cdot{\BF\Omega}^3_c-\bJ_3\cdot{\BF\Omega}^2_c\right]\label{Agninv}\\
\delta_{\lambda^2}S&=&\frac{4\pi\Lambda}\kappa\int d^3x
\lambda^2\left[
\bJ_3\cdot{\BF\Omega}^1_c-\bJ_1\cdot{\BF\Omega}^3_c\right]\\
\delta_{\lambda^3}S&=&\frac{4\pi\Lambda}\kappa\int d^3x
\lambda^3\left[
\bJ_1\cdot{\BF\Omega}^2_c-\bJ_2\cdot{\BF\Omega}^1_c\right]\label{Bgninv}
\end{eqnarray}
Here the symbols ${\BF\Omega}^i_c$ have been introduced to remember
that in Eqs.~(\ref{Agninv}--\ref{Bgninv})
the transverse degrees of freedom of the fields ${\BF\Omega}^i$
have been fixed by
means of Eq.~\ceq{confromot}.

At this point we denote with the symbol $\bcas$ the gauge invariant
extension of the action $S$ and we try for it the ansatz:
\beq{\bcas=S+S^1_b+S^2_b}{ansatz}
where $S^1_b$ and $S^2_b$ contain respectively terms which are linear
and quadratic in the fields ${\BF\Omega}^i$:
\begin{eqnarray}
S^1_b&=&\nonumber\\
&&\frac\Lambda\kappa \int d^3x\Omega^1_\mu(x)
\left[\oint_{\Gamma_3} dx^\nu_3
\frac{(x-x_3)^\mu}{|x-x_3|^3}b_\nu^2(x_3)
-\oint_{\Gamma_2}dx^\nu_2
\frac{(x-x_2)^\mu}{|x-x_2|^3}b_\nu^3(x_2)\right]\nonumber\\
&+&\frac\Lambda\kappa \int d^3x\Omega^2_\mu(x)
\left[\oint_{\Gamma_1} dx^\nu_1
\frac{(x-x_1)^\mu}{|x-x_1|^3}b_\nu^3(x_1)
-\oint_{\Gamma_3}dx^\nu_3
\frac{(x-x_3)^\mu}{|x-x_3|^3}b_\nu^3(x_3)\right]\nonumber\\
&+&\frac\Lambda\kappa \int d^3x\Omega^3_\mu(x)
\left[\oint_{\Gamma_2} dx^\nu_2
\frac{(x-x_2)^\mu}{|x-x_2|^3}b_\nu^1(x_2)
-\oint_{\Gamma_1}dx^\nu_1
\frac{(x-x_1)^\mu}{|x-x_1|^3}b_\nu^2(x_1)\right]\nonumber\\
\label{couone}
\end{eqnarray}
\begin{eqnarray}
S^2_b&=&\frac\Lambda{4\pi\kappa}\int d^3x\partial^\mu\Omega^1_\mu(x)\int d^3y
\partial^\nu\Omega^2_\nu(y)\oint_{\Gamma_3}dx_3^\rho\frac1{|x-x_3|}
\partial^{x_3}_\rho\frac1{|y-x_3|}\nonumber\\
&-& \frac\Lambda{4\pi\kappa}\int d^3x\partial^\mu\Omega^1_\mu(x)\int d^3y
\partial^\nu\Omega^3_\nu(y)\oint_{\Gamma_2}dx_2^\rho\frac1{|x-x_2|}
\partial^{x_2}_\rho\frac1{|y-x_2|}\nonumber\\
&+&\frac\Lambda{4\pi\kappa}\int d^3x\partial^\mu\Omega^2_\mu(x)\int d^3y
\partial^\nu\Omega^3_\nu(y)\oint_{\Gamma_1}dx_1^\rho\frac1{|x-x_1|}
\partial^{x_1}_\rho\frac1{|y-x_1|}\nonumber\\
\label{coutwo}
\end{eqnarray}
Since we are interested in the computation of Wilson loop amplitudes,
 the expressions of $S_b^1$ and $S_b^2$
have been written directly for the special
case in which the currents $\bJ_i$ are given by Eq.~\ceq{currdef}. The
generalization to currents of general form is straightforward.
The action $\bcas$ of Eq.~\ceq{ansatz} defines what we call here
truncated topological field theory or TTFT.

It is now possible to check that
$S^1_b$ and $S^2_b$
satisfy the conditions listed
below:
\begin{description}
\item {i)\phantom{ii}} The variations of $S^1_b$ and $S^2_b$ under the
gauge transformations \ceq{gautraome} satisfy the relations:
\beq{\delta_{\lambda^i}(S^1_b+S^2_b)+\delta_{\lambda^i}S=0}{sbtra}
for $i=1,2,3$.
This condition guarantees the gauge invariance of the action $\bcas$.
\item {ii)\phantom{i}} The addition of the counterterms $S^1_b$ and $S^2_b$
to the action $S$ does not affect the
equations of motion \ceq{confromot} and (\ref{Aeom}--\ref{Ceom})
for what is concerning the transverse components of the fields
${\BF\Omega}^i$ and $\tilde{\BF\Omega}^i$.
\item {iii)} $S_b^1$ and $S^2_b$
consists of topological terms, so that the
topological properties of the action $\bcas$ are not spoiled.
\end{description}
%
%
To verify the validity of Eq.~\ceq{sbtra}, it is sufficient to compute
the variations
$\delta_{\lambda^i}(S^1_b+S^2_b)$.
A straightforward calculation yields the following result:
\begin{eqnarray}
\delta_{\lambda^1}(S^1_b+S^2_b)&=&\frac{4\pi\Lambda}\kappa
\left[\oint_{\Gamma_2}d\bx_2\cdot
\lambda^1(x_2){\BF\Omega}^3_c(x_2)-
\oint_{\Gamma_3}d\bx_3\cdot
\lambda^1(x_3){\BF\Omega}^2_c(x_3)\right]\nonumber\\
\delta_{\lambda^2}(S^1_b+S^2_b)
&=&\frac{4\pi\Lambda}\kappa\left[\oint_{\Gamma_3}d\bx_3\cdot
\lambda^2(x_3){\BF\Omega}^1_c(x_3)-
\oint_{\Gamma_1}d\bx_1\cdot
\lambda^2(x_1){\BF\Omega}^3_c(x_1)\right]\nonumber\\
\delta_{\lambda^3}(S^1_b+S^2_b)
&=&\frac{4\pi\Lambda}\kappa\left[\oint_{\Gamma_1}d\bx_1\cdot
\lambda^3(x_1){\BF\Omega}^2_\mu(x_1)-
\oint_{\Gamma_2}dx_2\cdot
\lambda^3(x_2){\BF\Omega}^1_c(x_2)\right]\nonumber\\
\label{Cmiforgauvar}
\end{eqnarray}
It is easy to realize that the right hand sides of
Eqs.~\ceq{Cmiforgauvar} coincide exactly, apart from
a sign, with the gauge variations $\delta_{\lambda^i}S$ of
Eqs.~(\ref{Agninv}--\ref{Bgninv}) if 
the external currents $\bJ_i$ are given by Eq.~\ceq{currdef}.
This proves condition i)  and thus the gauge invariance of the action
$\bcas$\footnotemark{}\footnotetext{Let us notice however that,
similarly to what happens in the usual Chern-Simons field theories,
the gauge invariance of $\bcas$ is realized only up to terms of the
kind $\int d^3x\nablab\left[{\BF\lambda}^i
\cdot(\nablab\times\tilde{\BF\Omega}^i)\right]$, which can be discarded
only if the theory is defined on manifolds without boundary.}.

At this point we note that $S^1_b$ and $S^2_b$ contain only the
longitudinal components ${\BF\Omega}^i_L$ of the fields
${\BF\Omega}^i$.
Due to this fact, condition ii) is automatically satisfied and,
moreover, it is possible to perform the following substitutions in
Eqs.~(\ref{couone}--\ref{coutwo}):
\beq{{\BF\Omega}^i(x)\equiv
{\BF\Omega}^i_L(x)=\nablab(\Omega^i(x)-\Omega^i(x_0))}{subone}
where the $\Omega^i(x)$ are singlevalued scalar fields defined by the
relations:
\beq{\Omega^i(x)-\Omega^i(x_0)=\int^{\sbx}_{\sbx_0}
d\bx^{i\prime}\cdot
{\BF\Omega}^i_L(x')}{omegaidef}
In terms of the $\Omega^i(x)$'s, $S^1_b$ and $S^2_b$ can be written in
a form which is explicitly
metric independent in agreement with condition iii):
\begin{eqnarray}
S^1_b&=&\frac{4\pi\Lambda}\kappa\left[\oint_{\Gamma_3}dx_3^\rho\Omega^1(x_3)
b^2_\rho(x_3)-\oint_{\Gamma_2}dx_2^\rho\Omega^1(x_2)
b^3_\rho(x_2)\right]\nonumber\\
&+&\frac{4\pi\Lambda}\kappa\left[\oint_{\Gamma_1}dx_1^\rho\Omega^2(x_1)
b^3_\rho(x_1)-\oint_{\Gamma_3}dx_3^\rho\Omega^2(x_3)
b^1_\rho(x_3)\right]\nonumber\\
&+&\frac{4\pi\Lambda}\kappa\left[\oint_{\Gamma_3}dx_3^\rho\Omega^1(x_3)
b^2_\rho(x_3)-\oint_{\Gamma_2}dx_2^\rho\Omega^1(x_2)
b^3_\rho(x_2)\right]\label{Cindmetfor}
\end{eqnarray}
\begin{eqnarray}
S^2_b&=&\frac{4\pi\Lambda}\kappa\left[
\oint_{\Gamma_3}dx_3^\rho\Omega^2(x_3)\partial_\rho\Omega^1(x_3)
-\oint_{\Gamma_2}dx_2^\rho\Omega^3(x_2)\partial_\rho\Omega^1(x_2)\right.\nonumber\\
&+&\left.\oint_{\Gamma_1}dx_1^\rho\Omega^3(x_1)\partial_\rho\Omega^2(x_1)\right]
\label{sbtmetind}
\end{eqnarray}
We stress the fact that in the above formulas the undesired presence
of the variable $\bx_0$ introduced in Eqs.~(\ref{subone}) and
\ceq{omegaidef} has disappeared.

Before concluding this Section, let us show that the TTFT of
 Eq.~\ceq{ansatz}
is free from the inconsistences which
affected the naive BF--model with cubic interactions of
Eq.~\ceq{snaivevector}.
To this purpose, we need to study the classical equations of motion of
the fields. First of all, the variation of $\bcas$ with respect to
$\tilde{\BF\Omega}^i$ produces again the constraints \ceq{confromot}.
The solution of these equations has been already given in
Eq.~\ceq{consol}.
Varying instead $\bcas$ with respect to the
$\Omega^i_\mu(x)$'s 
one obtains the following relations:
\begin{eqnarray}
&&\frac\kappa{4\pi}\epsilon^{\mu\nu\rho}\partial_\nu
\tilde\Omega^1_\rho(x)+\Lambda
\epsilon^{\mu\nu\rho}\Omega^2_\nu(x)\Omega^3_\rho(x)\nonumber\\
&+&\frac\Lambda\kappa\left[\oint_{\Gamma_3}dx_3^\nu b^2_\nu(x_3)
\partial_x^\mu\frac1{|x-x_3|}-
\oint_{\Gamma_2}dx_2^\nu b^3_\nu(x_2)
\partial_x^\mu\frac1{|x-x_2|}\right]\nonumber\\
&+&\frac\Lambda{4\pi\kappa}\int
d^3y\partial^\nu\Omega^3_\nu(y)\oint_{\Gamma_2}dx_2^\rho\partial_x^\mu\frac
1{|x-x_2|} \partial_\rho^{x_2}\frac 1{|y-x_2|}\nonumber\\
&-&\frac\Lambda{4\pi\kappa}\int
d^3y\partial^\nu\Omega^2_\nu(y)\oint_{\Gamma_3}dx_3^\rho\partial_x^\mu\frac
1{|x-x_3|} \partial_\rho^{x_3}\frac 1{|y-x_3|}=0\label{corequone}
\end{eqnarray}
\begin{eqnarray}
&&\frac\kappa{4\pi}\epsilon^{\mu\nu\rho}\partial_\nu
\tilde\Omega^2_\rho(x)-\Lambda
\epsilon^{\mu\nu\rho}\Omega^1_\nu(x)\Omega^3_\rho(x)\nonumber\\
&+&\frac\Lambda\kappa\left[\oint_{\Gamma_1}dx_1^\nu b^3_\nu(x_1)
\partial_x^\mu\frac1{|x-x_1|}-
\oint_{\Gamma_3}dx_3^\nu b^1_\nu(x_3)
\partial_x^\mu\frac1{|x-x_3|}\right]\nonumber\\
&-&\frac\Lambda{4\pi\kappa}\int
d^3y\partial^\nu\Omega^1_\nu(y)\oint_{\Gamma_3}dx_3^\rho\frac
1{|y-x_3|} \partial_x^\mu\partial_\rho^{x_3}\frac 1{|x-x_3|}\nonumber\\
&-&\frac\Lambda{4\pi\kappa}\int
d^3y\partial^\nu\Omega^3_\nu(y)\oint_{\Gamma_1}dx_1^\rho\partial_x^\mu\frac
1{|x-x_1|} \partial_\rho^{x_1}\frac 1{|y-x_1|}=0\label{corequtwo}
\end{eqnarray}
\begin{eqnarray}
&&\frac\kappa{4\pi}\epsilon^{\mu\nu\rho}\partial_\nu
\tilde\Omega^3_\rho(x)-\Lambda
\epsilon^{\mu\nu\rho}\Omega^1_\nu(x)\Omega^2_\rho(x)\nonumber\\
&+&\frac\Lambda\kappa\left[\oint_{\Gamma_2}dx_2^\nu b^1_\nu(x_2)
\partial_x^\mu\frac1{|x-x_2|}-
\oint_{\Gamma_1}dx_1^\nu b^2_\nu(x_1)
\partial_x^\mu\frac1{|x-x_1|}\right]\nonumber\\
&+&\frac\Lambda{4\pi\kappa}\int
d^3y\partial^\nu\Omega^1_\nu(y)\oint_{\Gamma_2}dx_2^\rho\frac
1{|y-x_2|} \partial_x^\mu\partial_\rho^{x_2}\frac 1{|x-x_2|}\nonumber\\
&-&\frac\Lambda{4\pi\kappa}\int
d^3y\partial^\nu\Omega^2_\nu(y)\oint_{\Gamma_1}dx_1^\rho\frac
1{|y-x_1|} \partial_x^\mu
\partial_\rho^{x_1}\frac 1{|x-x_1|}=0\label{corequthree}
\end{eqnarray}
We note that the differences between
Eqs.~(\ref{corequone}--\ref{corequthree}) and
Eqs.\ceq{Cmiforgauvar} are limited to purely longitudinal terms, so
that there is no effect
on the transverse
components of the fields, in agreement with condition ii).
Moreover,
the
longitudinal
components of the fields ${\BF\Omega}^i$ do not generate further
constraints, contrarily to what
happens
in the naive BF--model of the previous Section.
As a matter of fact, if
one applies the operator $\partial_\mu$ to both sides of
Eqs.~(\ref{corequone}--\ref{corequthree}), it is easy to
realize that the contributions 
coming from the cubic
interactions present in the action $S$
cancel exactly against the new contributions coming
from $S_b^1$ and $S^2_b$. The reason is that now only the transverse components
of the fields are physical, while the longitudinal components are
associated to gauge degrees of freedom and remain thus undetermined by
the equations of motion.
In this way, the extension of the gauge symmetry to include the
transformations \ceq{gautraome} has solved
 the consistency problems discussed in Section 2.
\section{The Wilson Loop Amplitudes of the TTFT}
Summarizing the results of the previous Section, the action $\bcas$
describes a well defined topological field theory coupled to a set of
Wilson loops. Since the inconsistencies of the original action $S$
have been eliminated by the introduction of the terms $S^1_b$ and
$S^2_b$, we are now ready to compute the Wilson loop amplitudes of the
TTFT, which are given by:
\beq{\langle W_1(\Gamma_1) W_2(\Gamma_2) W_3(\Gamma_3)
\rangle_b=\int\left[\prod_{i=1}^3{\cal D}{\BF\Omega}^i
{\cal D}\tilde{\BF\Omega}^i\right]e^{-i\sbcas}}{wlcorr}
Here the average with respect to the fields ${\BF\Omega}^i$ and
$\tilde{\BF\Omega}^i$ has been written with the symbol
$\langle\ldots\rangle_b$ to distinguish it from the analogous average
of Eq.~\ceq{wlcorr}, in which the fields' behavior is governed by the
action $S$. Eq.~\ceq{wlcorr} represents the most general correlator of
Wilson loops.
In the following, we suppose that none of the Wilson loop operators is
trivial, i.~e.
\beq{W_i(\Gamma_i)\ne 1\qquad\qquad\qquad i=1,2,3}{conwls}
This condition is useful to rule out simpler subcases
which
are not of particular interest in the present context.
As a matter of fact, it is easy to
see
that, in the computation of
 amplitudes of the kind $\langle W_i(\Gamma_i) W_j(\Gamma_j)
\rangle_b$, $1\le i\ne j\le 3$, the
contributions coming from the cubic interaction terms present in
$\bcas$ are irrelevant. Thus, the TTFT behaves as a standard abelian
BF--model and delivers as topological invariants only the Gauss
linking numbers of the set of trajectories $\Gamma_i$ and $\Gamma_j$.

Let us now come back to the evaluation of Eq.~\ceq{wlcorr} under the
assumption \ceq{conwls}.
To eliminate the gauge freedom with respect to the transformations 
\ceq{gftilfie} and \ceq{gautraome}, we choose the Lorentz gauge
fixing, in which the fields are purely transverse:
\beq{\nablab\cdot{\BF\Omega}^i=\nablab\cdot\tilde{\BF\Omega}^i=0}{lgcon}
An immediate consequence of the Lorentz gauge is that $S^1_b$ and $S_b^2$
vanish identically because they contain only the longitudinal
components of the fields.
Thus, the amplitude
\ceq{wlcorr} can be written as
follows\footnotemark{}\footnotetext{Since the gauge group is abelian,
the contribution of the ghost fields to the Wilson loop amplitude
\ceq{wlcorr} amounts to a trivial normalization constant which can be
omitted.}:
\beq{\langle W_1(\Gamma_1) W_2(\Gamma_2) W_3(\Gamma_3)
\rangle_b=\int\left[\prod_{i=1}^3{\cal D}{\BF\Omega}^i
{\cal D}\tilde{\BF\Omega}^i\right]e^{-i\sbcas_q}}{wlsimp}
where the ``quantum'' action $\bcas_q$ is given
by:
\beq{\bcas_q=S-iS_{gf}}{quaact}
$S_{gf}$ being the gauge fixing term\footnotemark{}\footnotetext{Let
us note that, as it happens in C-S field theories, the topological actions 
$\bcas$ and $S$ keep the complex factor $i$ in the Feynman path
integral even in spaces equipped with Euclidean metrics. For this
reason, in agreement with our conventions, the gauge fixing term
appears in Eq.~\ceq{quaact} with a $-i$ factor in front.}:
\beq{S_{gf}=\int d^3x\left(\nablab \varphi^i\cdot{\BF\Omega}^i+
\nablab \tilde \varphi^i \cdot\tilde{\BF\Omega}^i\right)}{gaufixact}
In the above equation we have introduced
 the scalar fields $\varphi^i,\tilde\varphi^i$, which are Lagrange
multipliers imposing the gauge constraint \ceq{lgcon}.

We stress the fact that the disastrous effects caused by the
longitudinal components of the fields if one uses the naive action $S$
are brilliantly removed by the presence of the gauge fixing term
$S_{gf}$.
Indeed, one may easily check that the classical equations of motion
coming from the action \ceq{quaact} are free of inconsistences and 
admit the non-trivial solutions given below:
\begin{eqnarray}
{\BF\Omega}^i(x)&=&\bb^i(x)\label{Asolclaeqn}\\
\tilde{\BF\Omega}^i(x)&=&\frac\Lambda{2\kappa}\epsilon^{ijk}\int d^3y
\left(\nablab\frac 1{|x-y|}\right)\times\left[
\bb^j(y)\times\bb^k(y)\right]\label{Bsolclaeqn}\\
\tilde\varphi^i(x)&=&0\label{Csolclaeqn}\\
\varphi^1(x)&=&\frac{i\Lambda}\kappa\int d^3y \frac
1{|x-y|}\left[\bb^2(y)\cdot\bJ^3(y) - \bb^3(y)\cdot\bJ^2(y)\right]
\label{Dsolclaeqn}\\
\varphi^2(x)&=&\frac{i\Lambda}\kappa\int d^3y \frac
1{|x-y|}\left[\bb^3(y)\cdot\bJ^1(y) - \bb^1(y)\cdot\bJ^3(y)\right]
\label{Esolclaeqn}\\
\varphi^3(x)&=&\frac{i\Lambda}\kappa\int d^3y \frac
1{|x-y|}\left[\bb^1(y)\cdot\bJ^2(y) - \bb^2(y)\cdot\bJ^1(y)\right]
\label{Fsolclaeqn}
\end{eqnarray}
At this point it is possible to compute the generic Wilson loop
amplitude \ceq{wlsimp}.
As in the case of the naive BF-model discussed in Section 2, there are
only two connected Feynman diagrams, which are represented in
Fig.~\ref{feyrul}. The path integrals
in Eq.~\ceq{wlsimp} may be easily evaluated  integrating first over
the $\tilde{\BF\Omega}^i$ fields and then exploiting the constraints
\ceq{confromot} obtained in this way to perform the integration over the
fields ${\BF\Omega}^i$. Alternatively, one can derive the analytic
expression of $\langle W_1(\Gamma_1) W_2(\Gamma_2) W_3(\Gamma_3)
\rangle_b$ by means of successive Gaussian integrations. In both cases
the result is:
\beq{\langle W_1(\Gamma_1) W_2(\Gamma_2) W_3(\Gamma_3)
\rangle_b=N\mbox{exp}\left[-i\Lambda\int
d^3x\bb^1(x)\cdot(\bb^2(x)\times\bb^3(x))\right]}{wlfin}
where the $\bb^i$'s have been defined in \ceq{consol} and $N$ is a
normalization constant given by:
\beq{N=\int\left[\prod_{i=1}^3{\cal D}{\BF\Omega}^i
{\cal D}\tilde{\BF\Omega}^i\right]e^{-i\int d^3x\frac\kappa{4\pi}
\epsilon^{\mu\nu\rho} \Omega_\mu^i\partial_\nu\tilde\Omega_\rho^i}}{normcon}
We recall that the symbols $\Gamma_i$ in Eq.~\ceq{wlfin}
denote an ensemble of closed, non-intersecting paths
$\gamma_i,\gamma_i',\ldots$:
$\Gamma_i=\gamma_i+\gamma_i'+\ldots$. Due to the linearity properties
of the exponent in the right hand side of Eq.~\ceq{wlfin}, however, it
is clear that the amplitude $\langle W_1(\Gamma_1) W_2(\Gamma_2) W_3(\Gamma_3)
\rangle_b$ can be decomposed into a product of correlation functions
of three Wilson loops. For this reason, it will be sufficient to
consider from now on only the fundamental three loop correlation
function $\langle W_1(\gamma_1) W_2(\gamma_2) W_3(\gamma_3)
\rangle_b$, putting $\Gamma_i=\gamma_i$, $i=1,2,3$ in Eq.~\ceq{wlfin}.
As an upshot, the TTFT \ceq{ansatz} contains in
practice a single topological invariant, which appears in the exponent
of the right hand side of Eq.~\ceq{wlfin} and it is given by:
\beq{{\cal H}=\frac 13\int
d^3x\epsilon^{\mu\nu\rho}\epsilon^{ijk}b_\mu^i(x)b_\nu^j(x)b_\rho^k(x)
}{tisymm} 

To conclude this Section, we study the topological term ${\cal
H}$. In the following, it will be
convenient to interpret
the vector fields $b_\mu^i(x)$ of Eq.~\ceq{consol} as magnetic
fields\cite{edwards}
\beq{\bb^i(x)=-\frac
1\kappa\oint_{\Gamma_i}d\bx_i\times\frac{(\bx-\bx_i)} {|x-x_i|^3}}{magfieint}
generated by the currents $\bj_i=-\frac 1\kappa\bJ_i$. Indeed, it is
possible to see that the
$b_\mu^i(x)$'s
satisfy
the relations:
\begin{eqnarray}
\nablab\times\bb^i=\bj_i&\qquad\qquad&\nablab\cdot\bb_i=0\nonumber\\
\nablab\times\ba^i=\bb_i&\qquad\qquad&\nablab\cdot\ba^i=0\label{emfieeqs}
\end{eqnarray}
where the $\ba^i$'s are their associated electromagnetic potentials:
\beq{\ba^i=-\frac 1\kappa\oint_{\Gamma_i}d\bx_i\frac
1{|x-x_i|}}{emvecpot}
Moreover, one can introduce
the multivalued magnetic potentials
$v^i(x)$:
\beq{v^i(x)-v^i(x_0)=-\int_{x_0}^x d\bx'\cdot\bb^i(x')}{elpot}
defined in such a way that $\bb^i(x)=-\nablab
v^i(x)$\footnotemark{}\footnotetext{Mathematically,
each $v^i(x)$ is the solid  angle under which the trajetory $\Gamma_i$
appears as seen from a point $x$ (see e.~g.\cite{klePI}, Ch. 16).}.

The most straightforward interpretation of ${\cal H}$ is that of an Hopf
invariant of the underlying gauge group $U(1)\otimes U(1)\otimes
U(1)\equiv[U(1)]^3$. To show that, we build the $[U(1)]^3$ group
element
\beq{g(x)=e^{-i\sum_{i=1}^3 v^i(x)}}{groele}
It is now easy to check that, apart from a proportionality factor,
${\cal H}$ has exactly the form of the desired Hopf term:
\beq{{\cal H}\propto\epsilon^{\mu\nu\rho}\int d^3x
\frac {\partial v^1(x)}{\partial x^\mu}
\frac {\partial v^2(x)}{\partial x^\nu}
\frac {\partial v^3(x)}{\partial x^\rho}
g^{-1}\frac{\partial g}{\partial v^1}
g^{-1}\frac{\partial g}{\partial v^2}
g^{-1}\frac{\partial g}{\partial v^3}}{firint}

Another  form of ${\cal H}$ may be derived
introducing the Pauli matrices $\sigma^i$ and the vector fields
\beq{b_\mu(x)=\sigma^ib^i_\mu(x)}{postwo}
The expression of ${\cal H}$ as a function of $b_\mu(x)$ becomes:
\beq{{\cal H}=\frac 16
\int d^3x
\mbox{Tr}\left[\bb(x)\cdot(\bb(x)\times\bb(x))\right]}{htlike}
Here the symbol Tr denotes trace over the Pauli matrices.
To go back to the original formulation of ${\cal H}$
given in Eq.~\ceq{tisymm} it is sufficient to
use the relation
$\mbox{Tr}[\sigma^i\sigma^j\sigma^k]=2\epsilon^{ijk}$.
Apparently, from the above equation
 ${\cal H}$ coincides with an Hopf term
for the group $SU(2)$, which is not a symmetry group of our theory,
but of course one should remember that
$\bb(x)$ is not a pure $SU(2)$ gauge field configuration.

Finally, one can give a physical meaning
to ${\cal H}$ exploiting the electromagnetic analogy established by
Eqs.~(\ref{magfieint}--\ref{emvecpot})
and the fact that the $b_\mu^i(x)$'s
satisfy the classical equations of motion
(\ref{corequone}--\ref{corequthree}) in the Lorentz gauge. After some
calculations one finds:
\beq{{\cal H}=-\frac{4\pi}{3\kappa}\epsilon^{ijk}
\int\!\!\!\int_{\Sigma_i}
d\bS^i\cdot(\bb^j\times\bb^k)
}{forint}
Here
$\Sigma_i$ denotes an arbitrary surface whose
boundary is given by the contour
$\Gamma_i$, while $d\bS^i$ is the projection of the infinitesimal area
element of $\Sigma_i$ along the normal direction with respect to the
surface.
From Eq.~\ceq{forint} it turns out that ${\cal H}$ measures the sum
for $i=1,2,3$ of the fluxes of the vector fields
$\epsilon^{ijk}\bb^j\times\bb^k$ through the surfaces $\Sigma_i$.

\section{Conclusions}
In this work a new topological field theory has been constructed, the
TTFT of Eq.~\ceq{ansatz}, with the property that its perturbative
series contains only the finite set of Feynman diagrams given in
Fig.~\ref{feyrul}. The TTFT is exactly solvable and, besides the Gauss
link invariant which already appears in
abelian C-S field theories, it produces the further topological
invariant ${\cal H}$ of Eq.~\ceq{tisymm}. The latter has been
interpreted as an Hopf term in Eq.~\ceq{firint}. Another form of
${\cal H}$ has been given in \ceq{htlike}.
This equation suggests also an interesting generalization of
the TTFT, consisting in the replacement
of the cubic interaction present  in the naive action $S$ 
with a new interaction of the kind $\int d^3x
f^{ijk}{\BF\Omega}^i\cdot({\BF\Omega}^j\times{\BF\Omega}^k)$, where
$f^{ijk}$ denotes
the structure constants of a compact Lie group.

As a final remark, let us note
 that the derivation of the action of the TTFT starting
from the naive BF--model of Eq.~\ceq{snaivevector}
 has
some analogies with the way in which gauge invariance is implemented 
 in C-S based models of the quantum Hall effect\cite{qhe}.
In our case a fictitious one-dimensional ``boundary'', which lies on
the trajectories
$\Gamma_i$, appears due to the introduction of
the Wilson loops.
 The inconsistent constraints 
(\ref{Aextcon}--\ref{Cextcon}) arising in the naive BF--model are all
concentrated along these trajectories because of the particular form of the
currents $\bJ_i$ defined in Eq.~\ceq{currdef}.
The analogue of the edge state action of the quantum Hall effect is
given here by the boundary terms 
$S_b^1$ and $S_b^2$, which restore gauge invariance in the fields
${\BF\Omega}^i$ and eliminate in this way the inconsistences of the action
\ceq{snaivevector}. 


\begin{thebibliography}{99}
\bibitem{nemliq} M.~J.~Bowick, L.~Chander, E.~A.~Schiff and
A.~M.~Srivastava, {\it Science} {\bf 263} (1994), 943.
\bibitem{supflu} C. B\"auerle et al., {\it Nature} {\bf 382} (1996),
332;
V. M. H. Ruutu et al., {\it Nature} {\bf 382} (1996), 334.
\bibitem{wasserman} E. Wasserman, {\it Jour. Am. Chem. Soc.} {\bf 82}
(1960), 4433.
\bibitem{quahaleff} D. R. Leadly et al., {\it Phys. Rev. Lett.} {\bf
79} (1997), 4246.
\bibitem{chesim} 
A.~S.~Schwarz,
{\it Lett.\ Math.\ Phys.\ }  {\bf 2} (1978), 247;
S.~Deser, R.~Jackiw and S.~Templeton,
{\it Phys.\ Rev.\ Lett.\ } {\bf 48} (1982), 975;
S.~Deser, R.~Jackiw and S.~Templeton,
{\it Annals Phys.\ }  {\bf 140} (1982), 372;
C.~R.~Hagen,
{\it Annals Phys.\ }  {\bf 157} (1984), 342.
\bibitem{bfmod} G. Horowitz, {\it Comm. Math. Phys.} {\bf 125} (1989),
417;
A. S. Schwartz, {\it Comm. Math. Phys.} {\bf 67}(1979), 1;
M. Blau and G. Thompson, {\it Ann. Phys.} (NY) {\bf 205} (1991), 130.
\bibitem{edwards}
S. F. Edwards, {\it Proc.~Phys.~ Soc.}
{\bf 91} (1967), 513;
J.~Phys.~{\bf A1} (1968), 15.
\bibitem{klePI}
H. Kleinert,
{\em Path Integrals}, (World Scientific Publishing, 2nd Ed.,
Singapore, 1995).
\bibitem{qhe}
F.~Wilczek,
{\it Fractional Statistics and Anyon
Superconductivity},
(World Scientific, New Jersey 1990), Ch. 9;
X. G. Wen, {\it Phys. Rev.} {\bf B43} (1991), 11025.
\end{thebibliography}
\end{document}